\documentclass{cambridge6A} 
\pdfoutput=1

\usepackage[english]{babel}
\usepackage[includeheadfoot, width=442pt, height=720pt]{geometry}
\usepackage{amsmath}
\usepackage{amsfonts}
\usepackage{amssymb}
\usepackage{bbm,bm}
\usepackage{mathtools}
\usepackage{slashed}
\usepackage{mathalfa}
\usepackage{graphicx,caption,subfigure}
\usepackage{lmodern}
\usepackage{booktabs}
\usepackage[utf8]{inputenc}

\usepackage{cite}
\usepackage[normalem]{ulem}

\usepackage{xcolor}
\usepackage[colorlinks=false,urlbordercolor=red]{hyperref}
\usepackage{tensor}

\counterwithout{section}{chapter}
\counterwithout{equation}{chapter}

\begin{document}

\vspace*{-1.5cm}
\begin{flushright}
  {\small
  LMU-ASC 18/24\\
  MPP-2024-206
  }
\end{flushright}

\vspace{1.5cm}
\begin{center}
  {\LARGE \bf String dualities and modular symmetries\\[.3cm] in supergravity: a review} 
\vspace{0.65cm}

\end{center}

\begin{center}
{\Large
Niccol\`o Cribiori$^{a,b}$ and Dieter L\"ust$^{a,c}$
}
\end{center}

\vspace{0.1cm}
\begin{center} 
\emph{
$^a$Max-Planck-Institut f\"ur Physik (Werner-Heisenberg-Institut), \\[.1cm] 
   Boltzmannstra\ss e 8 ,  85748 Garching, Germany, 
   \\[0.1cm] 
 \vspace{0.3cm}
 $^b$ KU Leuven, Institute for Theoretical Physics,  \\[.1cm] 
        Celestijnenlaan 200D, B-3001 Leuven, Belgium, 
   \\[0.1cm] 
 \vspace{0.3cm}
$^c$ Arnold-Sommerfeld-Center for Theoretical Physics,\\ Ludwig-Maximilians-Universit\"at, 80333 M\"unchen, Germany \\[.1cm] 
    } 
\end{center} 

\vspace{0.5cm}

\begin{abstract}

We review the remarkable interplay between modular symmetries and supergravity, which has led to major advances in both physics and mathematics in recent decades. Our focus will be on four-dimensional models with $\mathcal{N}=1$ and $\mathcal{N}=2$ local supersymmetry. We will look at the early articles on the topic, but also touch on recent developments. These results testify to how supergravity, when supplemented with the appropriate assumptions, can be turned into a unique window into quantum gravity. 

\vspace{1.5cm}\begin{center}
{\it Invited contribution to the volume on ``Half a century of Supergravity"}
\end{center}

\end{abstract}

\thispagestyle{empty}
\clearpage

\section{Introduction}

That invariance under local supersymmetry implies general relativity, and therefore gravity, is a remarkable coincidence, which has provided us with an invaluable tool: supergravity \cite{Freedman:1976xh, Deser:1976eh}. Giving due credit to all the applications and implications of supergravity is not possible in a limited space. Therefore, we will focus on one of its most exceptional properties, namely the fact that, when supplemented with appropriate assumptions, supergravity can be transformed into a unique window into quantum gravity. These assumptions are modular symmetries.

This review is not intended to be exhaustive. Instead, the focus will be on how the very first articles on the topic arose and how these findings have shaped the field and may shape it in the future. Finally, we will also briefly mention some of the most recent developments, demonstrating how the subject is still moving and fruitful after more than thirty years.

Historically, supergravity models endowed with modular symmetries were constructed starting from \cite{Ferrara:1989bc}. One of the motivation driving that work was that modular symmetries arise in various places in string theory and indeed contain what today are called T- and S-dualities, here reviewed in section \ref{sec:duality}. Since supergravity is the low energy limit of string theory from a target space perspective, it is natural to investigate modular symmetries therein. In fact, the 
appearance of S-duality was first proposed in \cite{Font:1990gx}, precisely in the context of four-dimensional $\mathcal{N}=1$ supergravity models endowed with modular symmetries, as reviewed in section \ref{sec:N=1}. 
Even if we will not elaborate much on it in the following, duality symmetries were also instrumental for the M-theory hypothesis, which sparkled the second superstring revolution. This is perhaps the period in which the interplay between modular symmetries and supergravity shined the most.
Indeed, the seminal work by Seiberg and Witten \cite{Seiberg:1994rs} triggered several development leading to major advances in various areas of physics and mathematics. 
For the first time, certain non-perturbative phenomena in quantum gravity could be understood very precisely, albeit in unrealistic settings with enough preserved supersymmetry. From this unique historical moment, we recall the early work \cite{Ceresole:1995jg,deWit:1995dmj} on dualities in $\mathcal{N}=2$ supergravity and in particular in four-dimensional $\mathcal{N}=2$ heterotic compactifications, which we review in section \ref{sec:N=2}.
Today, the interplay between supergravity and modular symmetries is still source of fruitful investigations in various areas of physics and mathematics. For example, it has recently played an important role in the swampland approach to quantum gravity and in relation to the species scale, as we review in section \ref{sec:today}, where other recent applications are briefly mentioned as well. Finally, in the appendix A we recall some minimal and basic elements about $\mathcal{N}=1$ and $\mathcal{N}=2$ supergravity in four-dimensions.

A general lesson one can learn from this topic is that the very structure of supergravity rhymes very well with the theory of modular forms. While we will review several examples of this fact in what follows, and many more can be found in the literature we have not covered, why this happens is ultimately unclear to us. We believe that this is a remarkable coincidence, making the topic worthy of continued investigation in the future.

\section{Duality symmetries in string theory}
\label{sec:duality}

The main motivation for looking at modular transformations in supergravity comes from string theory. Indeed, among the most studied string dualities we have the so-called T- and S-transformations, which generate the modular group PSL$(2,\mathbb{Z})$ and, from a target-space perspective, play an important role in low energy effective actions  \cite{Ferrara:1989bc,Font:1990gx,Cvetic:1991qm}. 
We devote this section to briefly recall basic facts about these transformations and to explain the different physical interpretations that the group PSL$(2,\mathbb{Z})$ can have in spacetime.

\subsection{T-duality}

T-duality type transformations act on the geometric background parameters of a given string compactification. The most famous but also the simplest example is the transformation that inverts the modulus (radius) $R$ of a circle,
\begin{equation}
\label{Rto1/R}
\text{ T-duality}:\qquad R\longleftrightarrow {\alpha'}/R\,  ,
\end{equation}
where $\alpha'$ is the Regge slope. 
This is a symmetry of the entire string spectrum which exchanges internal momentum modes, the Kaluza-Klein states, with winding modes. Indeed, for a circle compactification the spectrum is
\begin{equation}\label{kkwinding}
m^2=\frac{n^2}{ R^2}+\frac{k^2R^2}{ \alpha'^2}\,,
\end{equation}
with $n\in\mathbb{Z}$ the quantized Kaluza-Klein momentum and $k\in\mathbb{Z}$ the string winding number, and it is invariant under (\ref{Rto1/R}) combined with $n\leftrightarrow k$.

The simple transformation (\ref{Rto1/R}) can be generalized in presence of a higher-dimensional moduli space. For example, for compactification on a $D$-dimensional torus, the corresponding discrete T-duality group is given by SO$(D,D,\mathbb{Z})$. In heterotic compactifications, if additional Wilson line moduli along the  maximal torus of the gauge group are present, then the T-duality groups is enlarged to SO$(D+16,D,\mathbb{Z})$.

For our discussion on effective supergravity models in sections \ref{sec:N=1} and \ref{sec:N=2}, the discrete duality group SL$(2,\mathbb{Z})$ and its projective version PSL$(2,\mathbb{Z})={\rm SL}(2,\mathbb{Z})/\{\pm 1\}$, usually denoted modular group, are most relevant.\footnote{The distinction between PSL$(2,\mathbb{Z})$ and SL$(2,\mathbb{Z})$ has a physical meaning. The first of these groups typically acts on moduli, while the second is usually acting on functions of moduli. This will be made manifest by $\mathcal{N}=2$ supergravity where, as reviewed in section \ref{sec:N=2}, physical vector multiplets moduli are projective coordinates of a special K\"ahler manifold, while symplectic sections are functions of the moduli transforming in the fundamental representation of Sp$(2n_V+2)$.} They are the basic building blocks for the constructions to be reviewed in the following.
When specialized to T-duality, this group acts on a complex scalar field,  $T=R^2+iB$, with $R^2$ the volume of some internal space and $B$ a background axionic field. 
Two notable transformations are: discrete shifts of the axion, $B\rightarrow B+1$, acting as $T\rightarrow T+i$ and stringy small-large radius dualities, acting for $B=0$ as $R\rightarrow 1/R$ while for $B\neq 0$ as  $T\rightarrow 1/T$. As is well-known, these two transformations generate the entire discrete T-duality group 
\begin{equation}
\label{sl2T}
{\rm PSL}(2,\mathbb{Z})_T\, :\qquad T\,\rightarrow\,\frac{aT-ib}{ icT+d}\, ,\quad a,b,c,d\in\mathbb{Z}\, ,\quad ad-bc=1\, .
\end{equation}

The geometric T-duality transformations are believed to be exact at all orders in string perturbation theory. However they are non-perturbative from the world-sheet perspective, with coupling constant proportional to $\sqrt{\alpha'}/R$. 
This non-perturbative nature will also be manifest in section \ref{sec:N=1}, since invariance/covariance of the low-energy effective action under SL$(2,\mathbb{Z})_T$ will require contributions from word-sheet instantons of the form $\exp(-T/\alpha')$, which are exponentially suppressed for $T\rightarrow\infty$.

\subsection{S-duality}

To discuss S-duality type transformations, we introduce a complex scalar field, $S=e^{-2\phi}+ia$. 
Here, $\phi$ corresponds to the
massless string dilaton, whose vacuum expectation value sets the string coupling constant $g_s = e^\phi$. 
S-duality transformations are elements of 
\begin{equation}
{\rm PSL}(2,\mathbb{Z})_S\, :\qquad S\,\rightarrow\,\frac{aS-ib}{ icS+d}\, ,\quad a,b,c,d\in\mathbb{Z}\, ,\quad ad-bc=1\, .
\end{equation}
Since they contain the inversion of the string coupling constant, $g_s\rightarrow 1/g_s$, they are non-perturbative with respect to it. 
The appearance of this SL$(2,\mathbb{Z})_S$ strong-weak coupling duality in string theory and its consequences for the low energy effective action were suggested for the first time in \cite{Font:1990gx}. 
They can be understood as the string generalization of the strong-weak coupling duality of ${\cal N}=4$ supersymmetric field theories due to Montonen and Olive \cite{Montonen:1977sn,Witten:1978mh}. 
The spectrum of these theories contains electrically charged particles, solitonic magnetic monopoles and dyons all described by the mass formula 
\begin{equation}
\label{dyonMO}
m^2=v^2\left(q_e^2g^2+\frac{(4\pi)^2q_m^2}{g^2}\right)\, ,
\end{equation}
where $v$ is the vacuum-expectation-value of the Higgs field, while $q_e$ and $q_m$ are quantized electric and magnetic charges. 
Notice that the formula \eqref{dyonMO} is the analogue of the string spectrum \eqref{kkwinding}  where the role of quantized momenta and winding numbers is played by the electric and magnetic charges, while the world-sheet coupling $\sqrt{\alpha'}/R$ corresponds to the gauge coupling constant $g$.  
Since ${\cal N}=4$ super-Yang-Mills {(coupled to gravity)} is obtained as the low-energy effective theory in toroidal compactification of the heterotic string, it was argued in \cite{Font:1990gx} that S-duality occurs in heterotic string theories due to the existence of solitonic heterotic $NS5$-branes \cite{Duff:1987qa,Strominger:1990et} wrapped around five-cycles of the six-dimensional torus and dual to the fundamental heterotic  string. After compatifications to four dimensions, electrically charged states arise from wrapped strings, while magnetic monopoles come from wrapped $NS5$-branes.
This picture for heterotic S-duality was subsequently further established  by Schwarz and Sen \cite{Schwarz:1993vs,Schwarz:1993mg,Sen:1994fa}.

In \cite{Font:1990gx}, another piece of evidence for S-duality in string theory was mentioned, namely  that it could have an eleven-dimensional
origin when viewing the ten-dimensional dilaton as the breathing mode of a circle compactification of eleven-dimensional supergravity. 
S-duality as strong weak-coupling duality in type IIB superstrings and its geometric origin via M-theory compactifications on a two-torus was then shown concretely in the work of Schwarz \cite{Schwarz:1995dk,Schwarz:1995jq}. 
As explained there, in the type IIB superstring the group SL$(2,\mathbb{Z})_S$ is transforming fundamental $F1$-strings and solitonic $D1$-branes into each other, as well as $NS5$-branes into $D5$-branes.
Type IIB S-duality plays also an essential role in the formulation of the twelve-dimensional F-theory \cite{Vafa:1996xn},
which provides a non-perturbative completion of the type IIB superstring by utilizing an auxiliary torus along the eleventh and twelfth dimensions, with the type IIB dilaton as complex structure modulus.\\

Let us close this section which some additional comments.
First, we recall that T-duality and S-duality can be combined and intertwined to build an extended duality group called U-duality, which describes the non-perturbative discrete symmetries of M-theory compactifications, see $e.g.$~\cite{Obers:1998fb} for a comprehensive review.  
Besides, in string theory there are so-called string-string dualities that map a certain weakly coupled string theory to another strongly coupled one. Perturbative states in one theory are mapped to non-perturbative wrapped branes in the dual. A well-known example is the ${\cal N}=2$ string-string duality between the heterotic string on $K3\times T^2$ and the type IIA superstring on a Calabi-Yau threefold    \cite{Kachru:1995wm}, to which we will come back in section \ref{sec:N=2}.
In fact, the discovery of duality symmetries was key to the proposal of M-theory \cite{Townsend:1995kk,Witten:1995ex}, which marked the beginning of the so-called second superstring revolution in the year 1995.

\section{Modular symmetries in ${\cal N}=1$ effective actions}
\label{sec:N=1}

In this section, we review the construction of ${\cal N}=1$ supergravity effective actions in four dimensions endowed with a non-trivial action of the modular group.  
They arise as low energy limit of several string compactifications, such as heterotic or type II superstring vacua. Whenever working solely with chiral multiplets, the resulting effective actions will be invariant under the discrete duality group SL$(2,\mathbb{Z})$.\footnote{As we will discuss when dealing with $\mathcal{N}=2$ models in the next section, couplings involving vector fields are not in\-va\-riant but only covariant under modular transformations; equations of motions are instead invariant. These statements do not depend on the number of preserved supercharges \cite{Gaillard:1981rj}. }
This requirement links supergravity to the beautiful mathematical field of modular forms, as first discussed in \cite{Ferrara:1989bc}.
Indeed the framework is quite general and the group SL$(2,\mathbb{Z})$ may implement both T- and S-dualities. 
While we will employ the notation of \cite{Ferrara:1989bc}, where the chiral multiplet scalar field $T$ is a geometric modulus, this field $T$ could be replaced for example by the complex dilaton field $S$ without any modification in the mathematical formalism, as already noticed in \cite{Font:1990gx}.

\subsection{The modular superpotential}

Given an $\mathcal{N}=1$ supergravity effective action coupled to chiral multiplets, the kinetic term of the latter is governed by a K\"ahler metric $g_{i\bar\jmath}$, which is the second derivative of the K\"ahler potential,  $g_{i\bar\jmath}=\partial_i\partial_{\bar\jmath}K(\phi_i,\bar\phi_i)$. 
Let us concentrate on a single complex scalar field $T$ with K\"ahler potential and metric
\begin{equation}
\label{eq:KpotT}
K=-\kappa\log(T+\bar T)\, ,\qquad g_{T\bar T}=\frac{\kappa}{(T+\bar T)^2}\, ,
\end{equation}
where the integer parameter $\kappa$ depends on the chosen string compactification.   For example, for the heterotic string on a six-dimensional compact manifold, such as a six-torus and its ${\cal N}=1$ orbifolds, $\kappa=3$ is used when the real part of $T$ parametrizes a two-cycle volume and thus $e^{-K}=(T+\bar T)^3$ is the volume of the whole compact space; instead, $\kappa=1$ corresponds to the case in which the scalar field is the complex heterotic dilaton $S$.

With this choice of K\"ahler potential, the moduli space is ${\rm SU}(1,1)/{\rm U}(1)$ and the theory is invariant under SL$(2,\mathbb{Z})$-transformations of the form \eqref{sl2T}, since they act as  K\"ahler transformations,
\begin{equation}
\label{eq:kaehlertransf}
K(T+\bar T)\rightarrow K(T+\bar T) +\omega(T)+\bar\omega(\bar T)\quad{\rm with}\quad \omega=\kappa\log(icT+d)\, ,
\end{equation}
and the K\"ahler metric is unchanged. 
Hence, modular-invariance is implemented via K\"ahler invariance. 
This remains true, albeit in a more complicated manner, also when turning on a superpotential $W(T)$, which in string theory may originate from non-perturbative effects. Demanding SL$(2,\mathbb{Z})$-invariance poses severe restrictions on the allowed form of $W(T)$ and establishes a connection to the the theory of modular forms.
To proceed, we need to recall the general supergravity fact that the superpotential transforms under K\"ahler transformations in such a way that the combination $G=K + \log W \bar W$ is K\"ahler-invariant, see appendix A.1 for more details. Then, to ensure SL$(2,\mathbb{Z})$-invariance of the effective action it is sufficient to require invariance of $G$, which occurs if \eqref{eq:kaehlertransf} is supplemented by
\begin{equation}\label{modtransw}
W(T)\rightarrow W(T)e^{i\alpha-\omega(T)}= e^{i\alpha}\frac{W(T)}{ (icT+d)^\kappa}\, ,
\end{equation}
where $\alpha$ is a constant possibly depending on $a,b,c,d\in \mathbb{Z}$. 
This means that the superpotential $W(T)$ must be a modular function of weight $w=-\kappa$ and the $T$-independent phase $\alpha$ is called multiplier system.\footnote{We refer to \cite{DHoker:2022dxx} for a comprehensive review on modular forms and their application in string theory.}

It is interesting to compare the first derivatives of $G(T, \bar T)$ with that of $\log W(T)$. Since $G$ is modular-invariant, from the theory of modular forms we know that 
\begin{equation}
\label{eq:GTsugra}
G_T(T,\bar T)=-\frac{\kappa}{T+\bar T}+\partial_T \log W(T)\, 
\end{equation}
must be a modular form of weight $w=2$. To fulfill this property, the holomorphic term in $G_T$ must transform under SL$(2,\mathbb{Z})$ as 
\begin{equation}
\partial_T \log W(T)\rightarrow (icT+d)^2\partial_T \log W(T)-i\kappa c(icT+d)\, ,
\end{equation}
Due to the last piece, which is not SL$(2,\mathbb{Z})$-covariant but nevertheless needed to compensate the modular transformation of the non-holomorphic term in $G_T$, the holomorphic quantity  $\partial_T \log W(T)$ is not a modular form (it is called a quasi modular form). This lesson is important: either we preserve holomorphy and give up on modularity, or we keep working with strict modular forms at the cost of departing from holomorphic functions.  The non-holomorphic term $-\frac{\kappa}{T+\bar T}$ is usually called holomorphic anomaly.

Up to this point the discussion has been fairly general. To show that the class of supergravity models just reviewed is not empty, we now look at concrete examples. 
Since we continue working with a single modulus $T$ and with K\"ahler potential \eqref{eq:KpotT}, the simplest superpotential with the correct modular transformation properties \eqref{modtransw} and without singularities in the fundamental domain is expressed in terms of the Dedekind $\eta$-function 
\begin{equation}
\label{modsupo}
W(T)=\frac{1}{\eta(iT)^{2\kappa}}
=e^{\pi \kappa T/6}\left(1+2\kappa e^{-2\pi \kappa T}+\kappa (3+2\kappa )e^{-4\pi  \kappa T}+\dots\right)
\, .
\end{equation}
This is indeed of the type expected from non-perturbative string effects, as we will review in section \ref{sec:N=1app}. 
With this choice of superpotential, the derivative $G_T$ is given by the non-holomorphic Eisenstein series $\hat G_2$,
\begin{equation}
\label{eq:hG2}
G_T(T,\bar T)=\frac{\kappa}{2\pi}\hat G_2(T,\bar T)=\frac{\kappa}{ 2\pi}\lim_{\epsilon\to 0^+} \sum_{m,n}{'} \frac{1}{(mT+n)^{2}|mT+n|^{\epsilon}}\, ,
\end{equation}
(the prime superscript on sums indicates that the term $m=n=0$ is omitted) whereas $\partial_T \log W(T)$ coincides with the holomorphic but not SL$(2,\mathbb{Z})$-covariant Eisenstein series $G_2$,
\begin{equation}
\label{eq:G2hol}
\partial_T \log W(T)= {\kappa\over 2\pi}G_2(T)={\kappa\over 2\pi}\biggl(2\zeta(2)+2\sum_{m=1}^\infty\sum_{n=-\infty}^\infty(mT+n)^{-2}\biggr)\, .
\end{equation}
Then, the gravitino mass term is
\begin{equation}
m_{3/2}^2(T,\bar T)=e^G=\frac{1}{(T+\bar T)^\kappa}|\eta(iT)|^{-4\kappa}\, ,
\end{equation}
while the scalar potential takes the form
\begin{equation}
V(T,\bar T)=\frac{1}{(T+\bar T)^\kappa}|\eta(iT)|^{-4\kappa}\left(\frac{\kappa}{4\pi^2}(T+\bar T)^2|\hat G_2(T,\bar T)|^2-3\right)\, ,
\end{equation}
and we will discuss its associated vacuum structure in some more detail in section \ref{sec:N=1app}.

At this stage, we can appreciate a remarkable interplay between supergravity and the theory of modular forms. Differently from the Eisenstein series $G_{2k}$ with $k>1$, the case $k=2$ is special since the infinite sum $\sum_{m,n}{'} \frac{1}{(mT+n)^{2}}$ is not absolutely convergent and its value depends on how the regularization is performed. One choice is to cut off the sums in a symmetric way, which leads to the holomorphic but not modular expression \eqref{eq:G2hol}. Another choice is to introduce a regulator $\epsilon >0$, which gives the non-holomorphic but modular form \eqref{eq:hG2}. Crucially, a regularization preserving both holomorphy and modularity does not exist, for there are no modular forms of weight $w=2$.
This structure has a precise counterpart in supergravity. Indeed, we have just shown that the mathematical fact 
\begin{equation}
\hat G_2(T,\bar T)=G_2(T)-\frac{2\pi}{T+\bar T}\, 
\end{equation}
corresponds to the supergravity relation \eqref{eq:GTsugra} precisely because the non-holomorphic term plays the role of the K\"ahler connection $\sim \partial_T K$ (see also formula \eqref{eq:DWdef}). We notice again a deep interplay between modular and K\"ahler transformations.

This simple model can be generalized by considering the superpotential \cite{Cvetic:1991qm}
\begin{eqnarray}\label{modsupogen}
W(T)=\frac{H(T)}{ \eta(iT)^{2\kappa}}\, ,
\end{eqnarray}
where $H(T)$ is a rational function of the absolute modular-invariant $j(iT)$. To avoid singularities inside the fundamental domain, one has to take $H(T)$ of the form
\begin{equation}
H(T)=\left(\frac{G_6(iT)}{\eta(iT)^{12}}\right)^m\left(\frac{G_4(iT)}{\eta(iT)^{8}}\right)^n{\cal P}(j(iT)),
\end{equation}
or equivalently (up to a constant) 
\begin{equation}
H(T)=\left(j(iT)-1728)\right)^{m/2}j(iT)^{n/3}{\cal P}(j(iT))\, .
\end{equation}
Here, $m$ and $n$ are positive integers, ${\cal P}(j(iT))$ is a polynomial in $j$, while $G_4$ and $G_6$ are Eisenstein series of modular weight $w=4$ and $w=6$ and with zeroes at $T=1$ and $T=\rho=e^{i\pi/6}$ respectively. 
Without loss of generality, the polynomial $\cal P$ can be chosen in such a way that its zeroes are away from these two above points. 
These superpotentials always diverge in the limit ${\rm Re}T\rightarrow\infty$.

The vacuum structure depends on the profile of the function $H(T)$ and on its zeroes.
The fixed points $T=1,\rho$ are always extrema of the potential but not necessarily minima. When they are minima, they are supersymmetric. Hence, modular invariance allows for vacua in which non-perturbative effects stabilize the modulus $T$ without breaking supersymmetry.
For more information we refer to \cite{Cvetic:1991qm}.

\subsection{The gauge kinetic function}

While we will not enter much into the details of couplings involving $\mathcal{N}=1$ vector multiplets, it is instructive to discuss the modular properties of the gauge kinetic function $f(z)$, which is an holomorphic function of ${\cal N}=1$ chiral multiplets. 
This function determines the field-dependent gauge coupling constant $g_a$ and the associated axionic $\theta_a$-term of a given factor of the gauge group $\mathcal{G}= \prod \mathcal{G}_a$ in the low energy effective action,
\begin{equation}
\label{eq:gtheta}
\frac{1}{g_a^2}={\rm Re} f(z) ,\qquad \theta_a={\rm Im} f(z) .
\end{equation}
We will see that due to certain loop effects, which again reflect a kind of holomorphic anomaly, $g_a^{-2}$ and  $\theta_a$ receive non-holomorphic contributions which cannot be written as the real or imaginary part of an holomorphic function. Hence, the relations \eqref{eq:gtheta} are valid at tree-level but can be spoiled by loop corrections. 

Microscopically, the specific form of $f(z)$ depends on which kind of string compactification scheme is considered.
For example, at string tree-level the gauge kinetic functions of gauge groups associated to $D$-branes in type II superstrings depend on the volume moduli of the cycles wrapped by the branes.
For heterotic string compactifications, the tree-level $f(z)$ is instead given by the (complex) dilaton field, $f(S)=S$.

Let us consider one-loop corrections in heterotic string compactifications, which historically were the first setups in which modular properties of the gauge kinetic function were investigated.
Loop corrections to type II gauge couplings were later computed in \cite{Lust:2003ky}.
Heterotic one-loop contributions to $f(z)$ are encoded into a certain integral on the world-sheet torus \cite{Dixon:1990pc}. For orbifold compactifications, this integral can be computed explicitly and results in the moduli-dependent (threshold) expression 
\begin{equation}
\label{gaugekin}
\Delta_a(T,\bar T)=\frac{b_a}{2}\log\left( |\eta(iT)|^4(T+\bar T)\right)\, ,
\end{equation}
{which is to be added to the tree-level gauge coupling, giving $1/g_a^2={\rm Re}f (z)+\Delta_a$.}
These threshold corrections receive moduli-dependent contributions only from the ${\cal N}=2$ subsectors of the heterotic orbifold compactifications, and the constants $b_a$ are the $\beta$-function coefficients of the gauge and matter fields from these subsectors. The $\Delta_a(T,\bar T)$ play an importan role in the unification of
the gauge coupling constants in heterotic string compactifications \cite{Ibanez:1991zv}.
As one can check, the above expression is invariant under SL$(2,\mathbb{Z})_T$-transformations. 
The non-holomorphic part in \eqref{gaugekin} cannot be expressed as the real (nor imaginary) part of a holomorphic function and, as such, it does not respect the tree-level relation between $f(z)$ and $g_a$ as in \eqref{eq:gtheta}. The reason is that this part originates from a non-local contribution to the effective ${\cal N}=1$ supergravity action
related to a one-loop anomaly of the ${\rm SU}(1,1)/{\rm U}(1)$ sigma-model \cite{LopesCardoso:1991ifk,Derendinger:1991hq,Kaplunovsky:1994fg}.
The anomaly is an infrared effect arising from the integration of massless string excitations. 
Instead, the holomorphic part of $\Delta_a$, namely the one-loop gauge kinetic function, $f_a^{(1)}=b_a\log\eta(iT)^2$, comes from the integration over the massive string momentum and winding modes on the world-sheet. 
Notice that the mathematical structure of $\Delta_a$ is quite analogous to that of the function $G=K+\log W\bar W$ discussed before: it contains an holomorphic part,  $f_a^{(1)}$, which formally corresponds to $\log W$ and transforms non-covariantly under modular transformations, and in addition there is the non-holomorphic term, $b_a\log (T+\bar T)$, which corresponds to $K$ and ensures modular-invariance of the total expression.

The above result is further refined when one considers a possible one-loop mixing between the heterotic dilaton $S$ and the modulus $T$, via a generalized Green-Schwarz mechanism which cancels part of the sigma-model anomaly. 
As a result, the field $S$ transforms at the one-loop level non-trivially under PSL$(2,\mathbb{Z})_T$-transformations,
\begin{equation}\label{Strans}
S\rightarrow S+\delta_{GS}\log (icT+d)\, ,
\end{equation}
where $\delta_{GS}$ is the constant characterizing the string one-loop mixing diagram between $S$ and $T$.
By including this Green-Schwarz effect, the total effective gauge coupling becomes then \cite{Ibanez:1992hc}
\begin{equation}
\label{eq:gathreshold}
\frac{1}{g_a^2}=\frac{S+\bar S}{2}+\frac{b_a}{2}\log(T+\bar T)+\frac{b_a-\delta_{GS}}{2}\log|\eta(iT)|^4\, .
\end{equation}
This expression contains the string one-loop holomorphic gauge kinetic function,
\begin{equation}
f_a^{(1)}=(b_a-\delta_{GS})\log\eta(iT)^2\,,
\end{equation}
which transforms under SL$(2,\mathbb{Z})_T$-transformations as
\begin{equation}\label{f1trans}
f_a^{(1)}\rightarrow f_a^{(1)}+(b_a-\delta_{GS})\log(icT+d)\, ,
\end{equation}
in such a way that the whole coupling $\frac{1}{g_a^2}$  is SL$(2,\mathbb{Z})_T$-invariant. 
For $b_a=\delta_{GS}$, $i.e.$ if the sigma-model anomaly is completely canceled by the Green-Schwarz mechanism, the one-loop holomorphic term proportional to $\log\eta(iT)^2$ is absent.

The integral representation of the threshold corrections $\Delta_a$ admits a suggestive interpretation in terms of the topological free energy $F$ of a two-dimensional torus compactification.\footnote{The topological free energy captures the contribution to the total free energy coming purely from massive states associated to the topology of the compactification. It is blind to those modes which are present also in the decompactified theory, such as massive oscillator modes.} 
In field theory, $F$ is obtained by integrating out  at one-loop all massive (bosonic or fermionic) states, 
\begin{equation}
e^F \simeq \int\lbrack{\cal D}\phi\rbrack e^{-\frac12 \phi M^2_\phi\phi^\dagger}\, .
\end{equation}
In string theory, in close analogy to the one-loop torus integral of \cite{Dixon:1990pc}, we can write $F$ as the (regularized) sum over all  massive BPS momentum and winding modes of a two-dimensional torus compactification \cite{Derendinger:1991hq}
\begin{equation}
\label{eq:Ftop}
F=\sum_{m,n}{'}\log\frac{|m+inT|^2}{(T+\bar T)}=\log\left( |\eta(iT)|^4(T+\bar T)\right)\, .
\end{equation}
The same expression was also obtained in \cite{Bershadsky:1993ta} as the genus-one free energy, $F_1$, of the topological string.

Finally, let us mention that just as for the superpotential $W$ also the holomorphic one-loop gauge kinetic function can be augmented by a completely modular-invariant term, such as \cite{LopesCardoso:1994ik}
\begin{equation}
f_a^{(1)}=b_a\log\eta(iT)^2+N\log j(iT)\, ,
\end{equation}
for some constant $N$.
Since the function $j(iT)$ has a triple zero at $T=\rho$, $i.e.$ $j(iT)\simeq (T-\rho)^3$ for $T\simeq\rho$, the gauge kinetic function above logarithmically diverges. This singularity is due to a finite number of  states becoming massless at $T=\rho$.

\subsection{Applications}
\label{sec:N=1app}

After having discussed the main features of the most simple $\mathcal{N}=1$ supergravity effective actions endowed with modular transformations, we now review some of their applications.\footnote{One important aspect we will not cover concerns the consequences of modular invariance on the spectrum of supersymmetry breaking soft masses \cite{Ibanez:1992hc}.}

\subsubsection{Tree level superpotential and Yukawa couplings}

A considerable amount of information about Yukawa couplings in orbifold compactifications can be obtained by exploiting the constraints imposed by modular symmetries \cite{Lauer:1989ax,Lerche:1989cs,Chun:1989se,Ferrara:1989qb}.
Consider some complex massless matter fields $\phi^i$ with $T$-dependent K\"ahler potential and kinetic terms
\begin{equation}
 K(\phi,\bar \phi)=\frac{ \phi^i \bar \phi^{\bar \jmath}}{(T+\bar T)^{\kappa_i}}\, , \qquad {\cal L}_{kin}=\frac{\partial \phi^i\partial \bar \phi^{\bar \jmath}}{ (T+\bar T)^{\kappa_i}}\, .
\end{equation}
In order for their kinetic term to be invariant under SL$(2,\mathbb{Z})_T$-transformations, it follows that the fields $\phi_i$ have modular weight $w_i=-\kappa_i$ and thus transform under modular transformations as
\begin{equation}
\phi^i\rightarrow \frac{\phi^i}{(icT+d)^{\kappa_i}}\, .
\end{equation}
Consider then a cubic tree level superpotential among the matter fields of the form
\begin{equation}
W(T,\phi_i)=W_{ijk}(T)\phi^i\phi^j\phi^k\, .
\end{equation}
As discussed, due to local supersymmetry the superpotential is a modular form with weight $w=-\kappa$ (we are assuming the same K\"ahler potential for $T$ as in \eqref{eq:KpotT}). 
Let us consider $\kappa=3$, as it is the case for a large class of heterotic orbifold compactifications.
Matter fields $\phi^i$ in the untwisted sector of the orbifolds generically have modular wights $w_i=-1$. 
It then follows that the Yukawa couplings $W_{ijk}$ must have vanishing modular weight. Requiring the absence of zeroes and poles, and since the only entire modular form with zero modular weight is the constant, we deduce that Yukawa couplings in these setups are just constants. 
More precisely, the untwisted Yukawa couplings have no $T$-dependence and are not exponentially suppressed by world-sheet instantons. 
On the other hand, matter fields in the twisted sectors of orbifold compactifications have different modular weights and hence the $W_{ijk}$ possess a non-trivial dependence on these moduli. 
  Concretely they are then given in terms of characters $\chi_\alpha(T)$ which transform under representations of a certain
principal congruence subgroup $\bar\Gamma\in \Gamma=SL(2,\mathbb{Z})_T$ which is preserved by the action of the orbifold twist. 
For example, for the ${\mathbb Z}_3$-orbifold the relevant subgroup is $\bar\Gamma =\Gamma(3)$, the charged matter fields have modular weights
$w_i=-1/3$ and the $\chi_\alpha(T)$ are the three level-one characters of the $SU(3)$ affine algebra. 
They are suppressed exponentially as $\exp(-T)$, implying that the twisted Yukawa couplings vanish in the large volume limit of the orbifold,
since the fixed points, at which the twisted fields are located, become infinitely far away from each other.

A selection of more recent articles on modular flavour symmetries in orbifold compactifications is \cite{deAdelhartToorop:2011re,Feruglio:2017spp,Baur:2019kwi,Nilles:2020nnc,Nilles:2020kgo,Baur:2020yjl,Nilles:2021glx}. Here, the modular group SL(2,$\mathbb{Z})_T$ is embedded into the bigger eclectic flavour symmetries.
These embeddings have interesting  consequences for the flavour structure of the Standard Model matter fields as well as for CP violation.

\subsubsection{Non-perturbative superpotential and gaugino condensation}

A very important and also frequently considered application of modular symmetries in supergravity concerns moduli stabilization and the possibility of breaking supersymmetry via non-perturbative effects \cite{Font:1990nt,Nilles:1990jv,Ferrara:1990ei}. 
In heterotic compactifications, gaugino condensation in hidden gauge sectors constitutes the prototype example of a non-perturbative contribution to the superpotential. The condensate can be parametrized as 
{$\langle\lambda\lambda\rangle\sim\exp\left(-\frac{3}{b_ag_a^2}\right)$}, where $b_a$ is the $\beta$-function coefficient of the non-Abelian hidden sector gauge group and $g_a$ is the associated coupling constant.
This expression translates into a non-perturbative contribution to the superpotential, which depends on both the heterotic dilaton $S$ and the geometric modulus $T$ as
\begin{equation}
\label{eq:Wcond}
W_{\lambda\lambda}(S,T)=e^{-\frac{3}{b_a}f(S,T)}=e^{-\frac{3}{ b_a}\left(S+(b_a-\delta_{GS})\log\eta(iT)^2\right)}\, .
\end{equation}
Using the modular transformation properties of $S$ and of the one-loop gauge kinetic function as given \eqref{Strans} and \eqref{f1trans}, one can show that \eqref{eq:Wcond} transforms as a modular function with weight $w=-3$,
as required for heterotic string compactifications on a six-dimensional internal space.
We can further simplify the non-perturbative superpotential by assuming that $\delta_{GS}=0$, such that $W_{\lambda\lambda}(S,T)$ becomes
\begin{equation}
W_{\lambda\lambda}(S,T)={e^{-\frac{3}{b_a}S}\over \eta(iT)^6}\, .
\end{equation}
Since $\eta (iT)\simeq \exp(-\frac{\pi}{12}{\rm Re}T)$ for ${\rm Re}T \to \infty$, this superpotential diverges in the large volume limit, while it vanishes in the limit of weak gauge coupling, ${\rm Re}S\rightarrow\infty$.

We can generalize the above expression by replacing $e^{-\frac{3}{b_a}S}$ with a more general function $\Omega(S)$, giving
\begin{equation}
W_{\lambda\lambda}(S,T)=\frac{\Omega(S)}{ \eta(iT)^6}\, .
\end{equation}
Taking as K\"ahler potential
\begin{equation}
K(S,T)=-\log(S+\bar S)-3\log(T+\bar T)\, ,
\end{equation}
the associated scalar potential is then 
\begin{equation}
V(S,T)=\frac{|(S+\bar S)\Omega_S-\Omega |^2+3|\Omega|^2\left(\frac{(T+\bar T)^2}{ 4\pi^2}|\hat G_2(iT)|^2-1\right)}{ (S+\bar S)(T+\bar T)^3|\eta(iT)|^{12}}\, ,
\end{equation}
where $\Omega_S=\partial \Omega/\partial S$.
This potential possesses a very interesting vacuum structure \cite{Font:1990nt}. Let us start from the $T$-dependence. In the decompactification limit, ${\rm Re}T\rightarrow\infty$, $V(S,T)$ diverges since $(T+\bar T)^3|\eta(iT)|^{12}$ {vanishes exponentially or, equivalently, the gravitino mass term diverges exponentially. Notice that in the same limit the gaugino condensate diverges, since more and more Kaluza-Klein states become light and contribute (negatively) to $b_a$. Hence, there is a dynamical obstruction to the large volume limit and the theory is forced to remain compactified.} 
The orbifold points $T=1,\rho$, and their modular transformed copies, are extrema of the potential and do not break supersymmetry in the $T$-direction; this is because $\hat G_2$ vanishes (only) at these points. $T=\rho$ is a local maximum, while $T=1$ is a saddle. 
The local minima of the potential are located around $T=1.2$ and break supersymmetry, since the F-term related to $T$ is non-vanishing at this point; {this is because $\hat G_2$ is non-zero.} 
Hence, gaugino condensation can break local supersymmetry spontaneously in the $T$-direction and the gravitino mass is
\begin{equation}
m_{3/2}=e^{G/2}=\frac{|\Omega(S)|}{\sqrt{(S+\bar S)(T+\bar T)^3}|\eta(iT)|^6}\, .
\end{equation}
As discussed in \cite{Cvetic:1991qm}, the asymptotic behavior for large $T$ and also the vacuum structure can be different when multiplying the superpotential by a modular-invariant function $H(T)$.

In addition to the minimization of the potential with respect $T$, one has also to consider the vacuum structure with respect to $S$. Along this direction,  one can show that the potential is minimized if 
\begin{equation}\label{omegas}
(S+\bar S)\Omega_S-\Omega =0\, ,
\end{equation}
and the corresponding F-term is vanishing.
Assuming that \eqref{omegas} is satisfied, one can calculate that the minimized potential is negative and find non-supersymmetric  $AdS_4$ vacua.
Since such vacua are in contradiction with swampland conjectures \cite{Ooguri:2016pdq}, it would be interesting to understand them better, especially with respect to the anti-de Sitter distance conjecture \cite{Lust:2019zwm}
and the associated relation between the AdS cosmological constant
and a tower of states.
Notice that the situation is similar to the KKLT scenario in type IIB string theory (before the uplift) \cite{Kachru:2003aw}, where indeed one has gaugino condensation or D3-instantons; nevertheless the KKLT $AdS_4$ vacua are supersymmetric. It would be interesting to understand if and how these heterotic $AdS_4$ vacua can be uplifted to de Sitter.

Finally, let us briefly discuss how the dilaton equation \eqref{omegas} can be satisfied. For the simplest choice $\Omega(S)=e^{-\frac{3}{b_a}S}$ there is no solution to \eqref{omegas} for finite $S$, hence the potential has a run-away direction along $S$, which drives it to $V\rightarrow 0$ for ${\rm Re}S\rightarrow\infty$.
One way to stabilize the dilaton is to turn on a constant, discrete heterotic three-form field $H_3$, such that $\Omega(S)=e^{-{3\over b_a}S}+H_3$
A second option is to assume that there is also a non-perturbative S-duality symmetry in heterotic compactifications.
In this case, $\Omega(S)$ must be a modular function of modular weight $w=-1$. The simplest way to realize this is when $\Omega(S)$ is given by the Dedekind function as
\begin{equation}
\Omega(S)={1\over \eta(S)^2}\, .
\end{equation}
Now the scalar potential has minimum for finite $S$, but diverges in the limit ${\rm Re}S\rightarrow\infty$.

\section{Modular symmetries in heterotic ${\cal N}=2$ effective actions}
\label{sec:N=2}

In this section, we review the construction of the $\mathcal{N}=2$ effective action arising from compactification of the heterotic string to four dimensions. Following \cite{deWit:1995dmj}, we discuss the crucial role played by modular symmetries, which leave the equations of motion, but not the action, invariant. More details on the supergravity rules governing the model can be found in \cite{Ceresole:1995jg}. We do not aim at giving a complete review of how to construct $\mathcal{N}=2$ effective actions, but refer to the standard reference \cite{Andrianopoli:1996vr}. Some essential information is also reviewed in appendix A.2. 
Instead, we will focus on one particular aspect, which we believe to be deep and remarkable. This is the fact that (local) supersymmetry, combined with modular symmetries and information about singularities of gauge couplings, allow to determine certain interactions ($i.e.$ the prepotential) at all order in perturbation theory.\footnote{A similar strategy can also be pursued for gauge couplings in $\mathcal{N}$=1 supergravity, see $e.g.$~the review \cite{Louis:1996ya}.} 
This is an explicit example of how supergravity, if supplemented with the appropriate additional assumptions, is powerful enough to get precise information about quantum gravity.

\subsection{Classical couplings}

We consider compactifications of the heterotic string on $T^2 \times K3$ preserving $\mathcal{N}=2$ supersymmetry in four dimensions. At string tree level, $i.e.$ classically, the fact that the dilaton does not mix with the other scalar fields and the continuous Peccei-Quinn symmetry of its axionic component uniquely fixes the special K\"ahler moduli space
\begin{equation}
\label{eq:MSK}
\mathcal{M}_{SK} = \frac{{\rm SU}(1,1)}{{\rm U}(1)} \otimes \frac{{\rm SO}(2,n_V-1)}{{\rm SO}(2) \times {\rm SO}(n_V-1)}\, ,
\end{equation}
where the first factor is the universal dilaton, while $n_V$ is the number of vector fields in the theory. 
This space is described by a prepotential
\begin{equation}
F(X) = -\frac{X^1}{X^0}\left[X^2 X^3-\sum_{I=4}^{n_V} (X^I)^2\right]\, ,
\end{equation}
and it is customary to single out the three scalar fields
\begin{equation}
S = -i \frac{X^1}{X^0}, \qquad T = -i \frac{X^2}{X^0}, \qquad U=-i\frac{X^3}{X^0}.
\end{equation}
The remaining fields are divided into the moduli $\phi^i = -i X^{i+3}/X^0$ with $i=1,\dots,P$, and the non-moduli scalars $C^a=-i X^{a+P+3}/X^0$ with $a=1,\dots,n_V-P-3$. 
This notation is to make the interpretation in terms of modular transformations more transparent, but also to be in analogy with the previous sections.
The three scalars are: the dilaton $S=g_s^{-2}+ia$, the K\"ahler modulus of the torus, $T=2 (\sqrt G+iB)$, and the complex structure modulus of the torus $U=(\sqrt G-iG_{12})/G_{11}$, with $G_{ij}$ the torus metric. 
In toroidal compactifications, the $\phi^i$ are Wilson line deformations and, since the rank of the gauge group is bounded by the heterotic central charge, ${\rm rank}(\mathcal{G})\leq 24$, we have $P\leq 20$. 
Given the prepotential, the rules of $\mathcal{N}=2$ supergravity allow to fix directly all couplings involving the vector multiplets sector (of the ungauged theory). For example, the K\"ahler potential is
\begin{equation}
K= -\log\left((S+\bar S)(T+\bar T)(U+\bar U)-\sum_i (\phi^i + \bar \phi^{\bar\imath})^2-\sum_a (C^a + \bar C^{\bar a})^2\right).
\end{equation}

To make the modular symmetries manifest, it is useful to perform a symplectic transformation on the setup just introduced. 
Physically, this is motivated by the fact that in the symplectic basis adopted above the vector field strength of the dilaton multiplet is strongly coupled in the regime of computational control, $i.e.$ for large ${\rm Re}S$. It is then convenient to replace it with the dual field strength, which is instead weakly coupled in the same limit. 
To this purpose, we perform a symplectic transformation 
\begin{equation}
\left(
\begin{array}{c}
X^\Lambda \\
F_\Lambda
\end{array}
\right)
\to 
\left(
\begin{array}{c}
\tilde{X}^\Lambda\\
\tilde{F}_\Lambda
\end{array}
\right),
\end{equation}
with $\tilde X^1 = F_1$,  $\tilde F_1 = -X^1$, $\tilde X^{\Lambda \neq 1} = X^{\Lambda \neq 1}$ and  $\tilde F_{\Lambda \neq 1} = F_{\Lambda \neq 1}$. This is a particular case of the general Sp$(2n_V+2, \mathbb{R})$ transformation  \eqref{eq:symplrotO} with only non-zero elements ${A^\Lambda}_\Sigma = {D_\Sigma}^\Lambda = \delta^\Lambda_\Sigma$, with $\Lambda, \Sigma \neq 1$, $B^{11}=1$, $C_{11}=-1$. 
In this new basis, the prepotential does not exist \cite{Ceresole:1995jg}. We will work in this basis and omit the tilde for convenience.

After the symplectic rotation, the K\"ahler potential and the gauge kinetic function read
\begin{align}
K &= -\log(S+\bar S) - \log(2z^\Lambda \eta_{\Lambda \Sigma}z^\Sigma),\\
\mathcal{N}_{\Lambda \Sigma} &= -2i \bar S \eta_{\Lambda \Sigma} + 2i (S+\bar S)\frac{\eta_{\Lambda \Gamma}\eta_{\Sigma \Delta}(z^\Gamma \bar{z}^\Delta+\bar{z}^{\Gamma} z^\Delta)}{z^\Gamma \eta_{\Gamma \Delta}{\bar z}^\Delta},
\end{align}
where  $z^\Lambda = X^\Lambda/X^0$, while $\eta_{\Lambda \Delta}$ is such that $A^T \eta A = \eta$. One can thus see that the gauge couplings $1/g^2 \sim {\rm Im} \mathcal{N}_{\Lambda \Sigma}$ vanish in the large ${\rm Re} S$ limit, $i.e.$ at weak string coupling. This basis has the advantage that the symmetry group SO$(2,2+P, \mathbb{R})$ of target space dualities is manifest and it acts as part of the Sp$(2n_V+2, \mathbb{R})$-rotations (see again formula \eqref{eq:symplrotO})
\begin{equation}
\label{eq:dualinvN=2classic}
X^\Lambda \to  {A^\Lambda}_\Sigma X^\Sigma, \qquad F_{\Lambda}\to {(A^{-1})^\Lambda}_\Sigma F^\Sigma.
\end{equation}
With respect to the nomenclature used in the appendix A.2, these are duality transformations of the first kind. 
The group SO$(2,2+P,\mathbb{R})$ is broken quantum-mechanically to SO$(2,2+P,\mathbb{Z})$, which contains as subgroup the product of the $T$- and $U$-modular groups, PSL$(2,\mathbb{Z})_T\times {\rm PSL}(2,\mathbb{Z})_U$. This acts on the geometric moduli and on the non-moduli scalars as 
\begin{align}
T  & \rightarrow \frac{aT-ib}{icT+d}\, ,\\
U  & \rightarrow  U - \frac{ic}{icT+d}\phi^i\phi^i\, ,\\
\phi^i  & \rightarrow \frac{\phi^i}{icT+d}\, ,\\
C^a  &\rightarrow \frac{C^a}{icT+d}\, ,
\end{align}
and we see that for non-vanishing $\phi^i$ the $T$- and $U$-directions are mixed. The above modular transformations are a subgroup of the Sp$(2n_V+2, \mathbb{R})$-transformations \eqref{eq:symplrotO}, restricted to integer coefficients and with
\begin{equation}
A=\left(\begin{array}{ccccc}
d & 0 & c & 0 & 0\\
0 & a & 0 & -b & 0\\
b & 0 & a & 0 & 0\\
0 & -c & 0 & d & 0\\
0 & 0 & 0 & 0 & 1_{n_V-3}
\end{array}
\right), \qquad D^{-1}=A^T, \qquad B=C=0.
\end{equation}
Besides the geometric moduli and the non-moduli scalars, the universal dilaton factor in \eqref{eq:MSK} has duality group $PSL(2,\mathbb{Z})_S$ acting as
\begin{equation}
S \to \frac{aS -ib}{icS+d},
\end{equation}
while it leaves the remaining scalar fields invariant.
According to the nomenclature in appendix A.2, this is a duality transformation of the third kind, and it is a subcase of \eqref{eq:symplrotO} with ${A^\Lambda}_\Sigma = d \delta^\Lambda_\Sigma$, ${D_\Sigma}^\Lambda = a\, \delta^\Lambda_\Sigma$, $C_{\Lambda \Sigma} = -2b \,\eta_{\Lambda\Sigma}$ , $B^{\Lambda \Sigma} = -\frac12 c \, \eta^{\Lambda \Sigma}$.

\subsection{Quantum corrections}

Having reviewed how modular symmetries act at the classical level, we consider now quantum corrections.  
In string perturbation theory, the couplings on the special K\"ahler moduli space of heterotic compactifications are modified beyond tree level, for the dilaton resides in a vector multiplet.\footnote{For the same reason, the couplings on the hypermultiplet moduli space are tree-level exact.} 
From the combination of holomorphy and axionic symmetries, we expect the general structure
\begin{equation}
F=F^{(0)}(S,\Phi)+ F^{(1)}(\Phi)+ F^{(NP)}(e^{-S},\Phi),
\end{equation}
where $F^{(1)}$ are string one-loop corrections, $F^{(NP)}$ are non-perturbative corrections, and we have denoted collectively with $\Phi$ all physical fields except the dilaton $S$.

We are not interested in $F^{(NP)}$ and thus we focus on $F^{(1)}$. Expanding at small $C^a$, we get 
\begin{equation}
F^{(1)}(\Phi)=-i(X^0)^2h^{(1)}(T,U,\phi^i)+\dots\, .
\end{equation}
and our goal is to determine $h^{(1)}(T,U,\phi^i)$. This cannot be an arbitrary function of the moduli, for it should respect any duality symmetry which is exact  at the quantum level. In our case, this symmetry is ${\rm SO}(2,2+P,\mathbb{Z})$ and it is part of the quantum (one-loop) corrected version of the  transformations \eqref{eq:dualinvN=2classic} which is
\begin{equation}
\label{eq:dualinvN=2quantum}
X^\Lambda \to  {A^\Lambda}_\Sigma X^\Sigma, \qquad F_{\Lambda}\to {(A^{-1})^\Lambda}_\Sigma F^\Sigma + C_{\Lambda \Sigma }X^\Sigma,
\end{equation}
where $D^{-1} = A^T$, $C = D\Lambda$, $\Lambda=\Lambda^T$. 
Classically $\Lambda=0$, but quantum-mechanically $\Lambda$ is a symmetric real matrix which is integer-valued in the appropriate symplectic base (not necessarily the one used now). 
With non-vanishing $\Lambda$ turned on, the transformations \eqref{eq:dualinvN=2quantum} are of the second kind, according to the nomenclature used in the appendix A.2.
These transformations generate discrete shifts in theta angles due to monodromies around semi-classical singularities in the moduli space, where massive string modes become massless. They act non-trivially on the one-loop term $F^{(1)}$, such that
\begin{equation}
\label{eq:h1transfX}
h^{(1)} \to h^{(1)} + \frac i2 \Lambda_{\Lambda \Sigma}X^\Lambda X^\Sigma/(X^0)^2 \, .
\end{equation}

To concretely determine $ h^{(1)} $, we now specify the analysis to the toroidal case with $\phi^i=0$. Here, the $T$-$U$ moduli space is locally
\begin{equation}
\mathcal{M}_{SK,TU} = \left(\frac{{\rm SO}(2,2)}{{\rm SO}(2) \times {\rm SO}(2)}\right)_{TU} = \left(\frac{{\rm SU}(1,1)}{{\rm U}(1)}\right)_T \otimes \left(\frac{{\rm SU}(1,1)}{{\rm  U}(1)}\right)_U
\end{equation}
and it has an associated duality group ${\rm PSL}(2,\mathbb{Z})_T \times {\rm PSL}(2,\mathbb{Z})_U$, acting as
\begin{align}
T &\to  \frac{aT-ib}{icT+d}\, ,\\
U  &\to  U\, ,\\
\label{eq:h1modtransf}
h^{(1)}(T,U)  &\to  \frac{h^{(1)}(T,U)+\Xi (T,U)}{(icT+d)^2}\, ,
\end{align}
for ${\rm PSL}(2,\mathbb{Z})_T$, and similarly for ${\rm PSL}(2,\mathbb{Z})_T$ upon exchanging $T\leftrightarrow U$. The form of $\Xi$ can be deduced by combining \eqref{eq:h1transfX} with \eqref{eq:h1modtransf}. 
For $\Xi=0$, $h^{(1)}$ is a modular form of weight $(w_T, w_U)=(-2,-2)$. Given that $\Xi$ is a quadratic polynomial, parametrizing the most general allowed ambiguities in the theta angles, even when $\Xi \neq 0$ it is still true that $\partial_T^3 \Xi = \partial_U^3 \Xi=0$. Since the $n$-th derivative of a modular form of weight $(1-n)$ is a modular form of weight $n+1$, we have that $\partial_T^3 h^{(1)}$ is a modular form of weight $(w_T, w_U)=(4,-2)$, while $\partial_U^3 h^{(1)}$  is a modular form of weight $(w_T, w_U)=(-2,4)$.

Since we are dealing with a modular form, we can exploit the fact that its analytic behavior is completely determined from the knowledge of its singularities and of its asymptotic regime, which for us corresponds to ${\rm Re}T\to \infty$ or ${\rm Re}U \to \infty$. We start from the singularities. These arise along the line $T =U$ (up to ${\rm PSL}(2,\mathbb{Z})$-transformations), where the gauge group is enhanced from $U(1)_T \times U(1)_U$ to $SU(2) \times U(1)$. Further enhancements occur when $T=U=1$ and $T=U= e^{i\pi/6}$. It turns out \cite{deWit:1995dmj} that the structure of the singularity is governed by threshold corrections such as \eqref{eq:gathreshold} with $b_a = \delta_{GS}$. Since this lesson is valid for both $T$ and $U$, and since $1/g_{TT}^2 \sim  {\rm Re} \partial_T^2 h^{(1)}$ and similarly for $U$, we get
\begin{equation}
h^{(1)}(T\sim U) \simeq \frac{1}{16\pi^2} (T-U)^2 \log (T-U)^2 + \text{regular}\, .
\end{equation}
The regular part can be fixed by looking at the asymptotic behavior of the modular form. Since for large $T$ or $U$ the gauge coupling cannot grow faster than a linear power of $T$ and $U$ (this is required for the theory to remain perturbative) and since the gauge coupling is related to the second derivatives of $h^{(1)}$ (this is dictated by $\mathcal{N}=2$ supersymmetry), we have that $\partial_{T,U}^3 h^{(1)}\to {\rm const}$ for ${\rm Re}T,{\rm Re}U \to \infty$. Combining all of the above information, we are led to the ansatz
\begin{equation}
\partial_T^3 h^{(1)} = \frac{f_{-2}(U)g_4(T)}{j(iT)-j(iU)}\, ,
\end{equation}
where $f_{-2}$ and $g_4$ are modular forms of weight $-2$ and $4$ respectively. They are constrained by the requirements that they should not have poles inside the fundamental domain, and that they should behave asymptotically as $f_{-2}(U)/j(iU) \to {\rm const}$, for ${\rm Re}U \to \infty$, and as $g_{4}(T)/j(iT) \to {\rm const}$, for ${\rm Re}T \to \infty$. This uniquely identifies $f_{-2} = E_4 E_6/\eta^{24}$ and $g_4 =E_4$, where $E_4$, $E_6$ are Eisenstein series. Hence, we arrived at
\begin{equation}
\partial_T^3 h^{(1)} =\frac{1}{2\pi}\frac{E_4(iT)E_4(iU)E_6(iU)}{\eta^{24}(iU)(j(iT)-j(iU)))} \, .
\end{equation}
Similarly, for $\partial_U^3 h^{(1)} $ we find
\begin{equation}
\partial_U^3 h^{(1)} =\frac{1}{2\pi}\frac{E_4(iU)E_4(iT)E_6(iT)}{\eta^{24}(iT)(j(iU)-j(iT)))}.
\end{equation}
One can also check that
\begin{equation}
\partial_T\partial_U h^{(1)} = -\frac{1}{4\pi^2}\log(j(iT)-j(iU))+\text{regular},
\end{equation}
which has the interesting property that the coefficient of the logarithmic divergence counts the number of string modes becoming massless \cite{LopesCardoso:1994ik}. 

The analysis just performed only determines the third derivatives of (the regular part of) $h^{(1)}$, because these are modular forms. In practice, this approach determines $h^{(1)}$ up to a quadratic polynomial $\Xi$. 
A complementary calculation of $h^{(1)}$ in string perturbation theory was performed by \cite{Harvey:1995fq}.

\section{Recent developments}
\label{sec:today}

So far we reviewed some of the major past developments in the construction of supergravity models endowed with modular symmetries. However, this is an active field of research which is still witnessing new advances. In this section, we briefly review one of these advances related to the notion of species scale in quantum gravity, and then just mention other recent applications.

\subsection{Modular-invariant species scale}

The species scale \cite{Veneziano:2001ah,Dvali:2007hz} gives an upper bound on the ultraviolet cutoff of gravitational effective theories. Recently, its dependence in terms of the moduli of certain string compactifications has been investigated \cite{vandeHeisteeg:2022btw,vandeHeisteeg:2023ubh,Cribiori:2023sch,vandeHeisteeg:2023dlw,Castellano:2023aum} and connected to black hole entropy \cite{Cribiori:2022nke} and thermodynamics \cite{Cribiori:2023ffn,Basile:2024dqq,Herraez:2024kux}. As we are going to review, it turns out that the moduli-dependent expression is a modular-invariant function, at least in setups with enough supersymmetry to have control over the computation. We will just focus on four-dimensional models and refer to \cite{vandeHeisteeg:2023dlw,Castellano:2023aum} for similar examples in higher dimensions.

A convenient way to introduce the species scale is as the inverse horizon radius of the smallest possible black hole in the theory; see e.g.~\cite{Bedroya:2024ubj} for a recent discussion. 
Let us consider once more the heterotic string compactified on $K3 \times T^2$, or the dual type IIA superstring compactified on $(K3 \times T^2)/\mathbb{Z}_2$. 
In this framework, BPS black holes are specified by the choice of a prepotential and of electric and magnetic charges. 
As for the latter, the central charge of the supersymmetry algebra gives the mass of BPS objects with electric and magnetic charges $(q_\Lambda, p^\Lambda)$ as
\begin{equation}
Z = e^K(q_\Lambda X^\Lambda - p^\Lambda F_\Lambda).
\end{equation}
As for the prepotential, in the type IIA frame we take it to be\footnote{Here and in the following $F_1$ is not to be confused with $\partial_1 F$. The latter will never appear explicitly in the discussion.}
\begin{equation}
    F=F_0+F_1 = -\frac16\frac{C_{ijk}X^i X^j X^k}{X^0} -\frac{1}{24}c_{2i}\frac{X^i}{X^0},
\end{equation}
and the heterotic expression can be obtained upon replacing $C_{ijk}\to C_{1jk}\delta_{i1}$ and $c_{2i}\to 24\delta_{i1}$; recall that $i,j=1,\dots,n_V$.
At the level of the spacetime effective action, $F_{1}$ couples to an higher-derivative $R^2$-correction \cite{LopesCardoso:1998tkj}, which descends from the known $R^4$-term in eleven dimensions \cite{Green:1997as,Antoniadis:1997eg}. The coupling arises via a graviphoton $\mathcal{N}=2$ chiral multiplet which we are implicitly setting to its constant background value; for details in this respect we refer to \cite{LopesCardoso:1998tkj}. In fact, $F_{1}$ is the first of a series of terms, $F_{g}$, which appear at genus-$g$ in superstring perturbation theory.\footnote{There is also a corresponding series of terms, $F^{(g)}$, in heterotic string perturbation theory. Since the heterotic dilaton is in a vector multiplet, while the type IIA dilaton is in a hypermultiplet, we have $e.g.$~$F^{(0)}=F_0^{(0)}+F_1^{(0)}+\dots$, $F^{(1)}=F_0^{(1)}+F_1^{(0)}+\dots$, while all $F_g$ are tree-level exact in type IIA perturbation theory.} 
They are related to genus-$g$ topological string amplitudes \cite{Antoniadis:1993ze,Antoniadis:1995zn,Antoniadis:1996qg}, but the relation can be subtle \cite{Cardoso:2014kwa}.

BPS black holes are always extremal and as such their entropy is captured by the attractor mechanism \cite{Ferrara:1996dd}. However, in the presence of higher-derivative corrections the Bekestein-Hawking formula is known to be incomplete, but one can use instead the Wald formula. 
In the present setup, this gives \cite{LopesCardoso:1998tkj}
\begin{equation}
\mathcal{S}_{BH} = \pi\left(Z\bar Z + \frac16 c_{2i}{\rm Im}\left(\frac{X^i}{X^0}\right)\right)\, ,
\end{equation}
where the second term comes genuinely from $F_{1}$.
The species scale can be found by minimizing this expression in terms of the charges. Since special geometry dictates that $Z \bar Z = e^{-K}$, the first term on the right hand side cannot be minimized at zero. Indeed, what happens is that the first and the second term, when minimized, are both quadratic in the charges. Thus, the entropy is effectively lower bounded by the second term. 
We find then \cite{Cribiori:2022nke}
\begin{equation}
\mathcal{S}_{BH} \gtrsim c_{2i}{\rm Im}\left(\frac{X^i}{X^0}\right) \simeq F_{1}\, ,
\end{equation}
(axions are vanishing in this black hole solution) which sets the species scale $\Lambda_{sp}$ as the inverse horizon radius of the black hole with entropy $F_{1}$, namely 
\begin{equation}
\label{eq:LambdaSpF1}
\Lambda_{sp} \simeq \frac{M_P}{\sqrt{F_{1}}}.
\end{equation}
In turn, this tells us that the dependence of the species scale on the $\mathcal{N}=2$ vector multiplet moduli is captured by the (topological) string one-loop correction $F_{1}$ \cite{vandeHeisteeg:2022btw}, which we have already encountered in \eqref{eq:Ftop}.

From supergravity we have access to the large volume, weak coupling expression $F_1\simeq c_{2i}(X^i/X^0)$, which is linear and holomorphic in the 2-cycle volume moduli $T^i$ (in type IIA frame), or in the dilaton $S$ (in heterotic frame), but not invariant under PSL$(2, \mathbb{Z})$-transformations. 
At the cost of relaxing holomorphy, we can construct a modular-invariant completion of the species scale \cite{Cribiori:2023sch}. 
Let us consider the heterotic frame as an example. We want to have a PSL$(2,\mathbb{Z})$-invariant expression with the same asymptotic behavior at zero and infinity. 
This requirement leads to \cite{LopesCardoso:1999cv,Cribiori:2023sch}
\begin{equation}
\label{eq:F1S}
F_1 \simeq -6 \log ((S+\bar S)|\eta(iS)|^4),
\end{equation}
and a similar expression holds in the type IIA frame, upon replacing $S \to T^i$; see once more formula \eqref{eq:Ftop} or equivalently  the (heterotic) threshold correction \eqref{eq:gathreshold}.

At the boundary of the moduli space, ${\rm Re} S \to \infty$, the Dedekind function reproduces the supergravity result \eqref{eq:LambdaSpF1} and the species scale approaches the string scale, $\Lambda_{sp} \simeq M_s$. 
However, the non-holomorphic term in \eqref{eq:F1S} is responsible for an additive logarithmic correction to the species scale, slightly increasing its tree-level value,
\begin{equation}
\Lambda_{sp} \simeq M_s\left(1+\frac32 ({\rm Re}S)^{-2}\log {{\rm Re}S}^2\right).
\end{equation}
This is a genuinely quantum gravitational effect, originating from the holomorphic anomaly.
Hence we showed that, combined with modular symmetries, supergravity allows us to obtain an expression of the species scale, and thus an upper bound on the ultraviolet cutoff of gravitational effective theories which is valid over the whole moduli space. This is yet another example of how powerful supergravity may be, when supplemented with the proper external input.

\subsection{Other applications}

Among the modern lines of research combining modular symmetries and supergravity, we would like to recall the one prompted by \cite{Feruglio:2017spp}, in which neutrino masses are postulated to be modular forms. Together with supersymmetry, this assumption leads to stringent constraints on neutrino masses and mixing angles which can be compatible with current experimental data, see $e.g.$~\cite{Criado:2018thu,Feruglio:2021dte,Feruglio:2022koo}.
Another interesting and recent development consists in studying the interplay of modular symmetries and supergravity for what concerns swampland conjectures \cite{Gonzalo:2018guu}, in search for or excluding de Sitter vacua  \cite{Leedom:2022zdm}, or to possibly construct inflationary models, see $e.g.$~\cite{Kobayashi:2016mzg,Casas:2024jbw,Kallosh:2024ymt,Antoniadis:2024ypf}.
On a more formal level, the combination of eleven-dimensional supergravity together with the Emergence Proposal \cite{Heidenreich:2017sim,Grimm:2018ohb} and the species scale, has recently allowed \cite{Blumenhagen:2024ydy} to reconstruct a non-trivial relation involving Eisenstein series, which was established in \cite{Bossard:2015foa,Bossard:2016hgy} via completely different methods. Even more recently, in \cite{Duff:2024ikg} it has been speculated on the possible existence of a novel Montonen-Olive duality which should explain the vanishing of the  $\beta$-function in gauged $\mathcal{N}> 4$ supergravity. Since Montonen-Olive duality can be considered among the precursors of the topic here reviewed, we believe that this is the right point to conclude by stressing how these, and many other developments we did not mention, are testifying how fruitful the interplay of supergravity with modular symmetries can be.

\paragraph{Acknowledgments.}
We thank A.~Ceresole and G.~Dall’Agata for kindly inviting us to contribute to this volume. The main part of the work presented in this review was performed in collaboration with many colleagues and in particular with M.~Cvetic, B~de Wit, S.~Ferrara, A.~Font, L.~Ibanez, V.~Kaplunovsky, W.~Lerche, G.~Lopes Cardoso, J.~Louis, T.~Mohaupt,  F.~Quevedo, S.~Theisen and N.~Warner.
We are very grateful to all of them for the very pleasant and fruitful collaborations. 
The work of D.L.~is supported by the Origins Excellence Cluster and by the German-Israel-Project (DIP) on Holography and the Swampland.

\appendix

\counterwithout{equation}{chapter}

\section*{A Elements of supergravity in four dimensions}

In this appendix we recall the basic ingredients of supergravity theories in four spacetime dimensions with four or eight preserved supercharges. For more details we refer to excellent textbooks such as \cite{Wess:1992cp,Freedman:2012zz,DallAgata:2021uvl}, and to \cite{Ceresole:1995ca,Andrianopoli:1996cm} for additional information on the rich structure of ${\cal N}=2$ supergravity.

\subsection*{A.1 ${\cal N}=1$ supergravity}
\label{sugra}
Four supercharges is the minimal amount of supersymmetry that can be preserved in four dimensions. 
The associated theory is denoted $\mathcal{N}$=1 supergravity and it is constructed out of the gravity multiplet, containing the graviton $g_{\mu\nu}$ and the gravitino, possibly interacting with matter.  
The matter couplings employed the most involve chiral multiplets, containing a complex scalar and a majorana (or Weyl) fermion,  $\{z,\chi\}$,  and vector multiplets, containing a fermion, called gaugino, and a real vector, $\{\lambda, A_\mu\}$.

The effective action is completely fixed by specifying the K\"ahler potential $K(z,z)$, the superpotential $W(z)$ and the gauge kinetic function $f(z)$. The first is a real function of the scalar fields, while the latter are holomorphic.
The bosonic sector of the theory reads 
\begin{equation}
\label{N=1LagBos}
\begin{aligned}
e^{-1}\mathcal{L} &=\frac12 R - g_{i\bar\jmath} D_\mu z^i D^\mu \bar z^{\bar \jmath} -\frac14 {\rm Re}f_{\Lambda \Sigma} F_{\mu\nu}^{\Lambda} F^{\mu\nu\,\Sigma} +\frac 18 {\rm Im}f_{\Lambda\Sigma} \epsilon^{\mu\nu\rho\sigma} F_{\mu\nu}^\Lambda  F_{\rho\sigma}^\Sigma -V,
\end{aligned}
\end{equation}
where indices $i,j=1,\dots,n_C$ are counting chiral multiplets, while $\Lambda,\Sigma=1,\dots,n_V$ vector multiplets. The covariant derivative on the scalar fields is $D_\mu z^i = \partial_\mu z^i - A^\Lambda_\mu k_\Lambda^i$, while on the superpotential is
\begin{equation}
\label{eq:DWdef}
D_i W = \partial_i W + W \partial_i K .
\end{equation}
The scalar potential is made up of two contributions, $V=V_F + V_D$, associated to the two classes of supersymmetry-breaking interactions, namely F-terms and D-terms. 
The F-term potential reads
\begin{equation}
V_F = e^K\left(g^{i\bar\jmath}D_{i}W\bar D_{\bar \jmath} \bar W-3 W \bar W\right), 
\end{equation}
while the D-term potential is
\begin{equation}
V_D = \frac12 ({\rm Re }f^{-1})^{\Lambda \Sigma} \mathcal{P}_\Lambda \mathcal{P}_\Sigma,
\end{equation}
with $\mathcal{P}_\Lambda(z,\bar z)$ real moment maps related to the Killing vectors $k_\Lambda^i = - ig^{i\bar\jmath}\partial_{\bar \jmath}\mathcal{P}_\Lambda$. Fayet-Iliopoulos terms arise as ambiguities in choosing $\mathcal{P}_\Lambda$ up to constants $\xi_\Lambda $.

Supersymmetry forces the scalar manifold to be K\"ahler-Hodge. We recall two consequences of this fact. First, $K$ and $W$ are not independent but related by K\"ahler transformations, $K\to K+\omega(z)+\bar \omega(z)$ and $W\to e^{-\omega}W$, with $\omega=\omega(z)$ an arbitrary holomorphic function. Hence, it is customary to introduce and work with the K\"ahler-invariant quantity
\begin{equation}\label{gfunction}
G=K+\log W \bar W,
\end{equation}
which is indeed the only combination of $K$ and $W$ appearing in the effective action.
Second, since the superpotential is not invariant under K\"ahler transformations, $\partial_i W$ is not covariant and one has to introduce \eqref{eq:DWdef}. As discussed in section \ref{sec:N=1}, this structure is precisely what is needed to have an analogy with the non-holomorphic Eisenstein series $\hat G_2$: the connection part in $D_i W$ plays the role of the non-holomorphic term. 
K\"ahler-invariance of the effective action is highly constraining the couplings and modular transformations can only be implemented as a special case of K\"ahler transformations. Then, modular-invariance works because of K\"ahler-invariance.

\subsection*{A.2 ${\cal N}=2$ supergravity}
\label{app:N=2review}

Eight preserved supercharges happens to be the right balance between too much and to few supersymmetry. The associated theory is denoted ${\cal N}=2$ supergravity and it is a remarkably rich and vast framework in which several questions about quantum gravity can be answered very precisely. 
Its main ingredients are the gravity multiplet and matter multiplets, among which we recall vector and hyper multiplets. The gravity multiplet contains the graviton, an SU$(2)_R$ doublet of gravitini and the graviphoton. Vector multiplets contains one complex scalar, an SU$(2)_R$ doublet of fermions and one vector, $\{z,\lambda^A, A_\mu\}$. Hyper multiplets contain four real scalars and two fermions. The scalar manifold has a product structure
\begin{equation}
\mathcal{M} = \mathcal{M}_{SK} \otimes \mathcal{M}_Q.
\end{equation}
The manifold $\mathcal{M}_{SK}$ is spanned by the $n_V$ scalars $z^i$; since it is special K\"ahler, it is endowed with a  Sp$((2n_V+2),\mathbb{R})$ vector bundle \cite{Craps:1997gp}. The manifold $\mathcal{M}_Q$ is spanned by quaternionic scalars. We will not deal with quaternionic geometry, but we refer the interested reader to \cite{Andrianopoli:1996cm} for more details.

The structure of the theory is deeply governed by symplectic transformations. Indeed, interactions on $\mathcal{M}_{SK}$ are fixed by specifying symplectic sections $(X^\Lambda, F_\Lambda)$, with $\Lambda=0,1,\dots, n_V$. Typically, one starts from a prepotential $F=F(X)$, such that $F_\Lambda = \partial_\Lambda F$. Crucially, the prepotential does not exist in all symplectic frames, while sections do.  
The physical scalars of the vector multiplets are recovered as projective coordinates on $\mathcal{M}_{SK}$, namely
\begin{equation}
\label{normcoord}
z^i = \frac{X^i}{X^0}, \qquad X^0 \equiv 1.
\end{equation}
Given the symplectic section, one can construct the K\"ahler potential
\begin{equation}
\label{N=2Kaelpot}
K = - \log i \left(\bar X^\Lambda F_\Lambda - X^\Lambda \bar F_\Lambda\right),
\end{equation}
the gauge kinetic matrix
\begin{equation}
\mathcal{N}_{\Lambda \Sigma} = \bar F_{\Lambda\Sigma} + 2i \frac{{\rm Im}( F_{\Lambda\Gamma}){\rm Im}( F_{\Sigma\Delta})X^\Gamma X^\Delta }{{\rm Im}( F_{\Gamma\Delta}) X^\Gamma X^\Delta},
\end{equation}
and in fact the whole vector multiplet sector of the theory, which reads
\begin{equation}
\begin{aligned}
e^{-1}\mathcal{L} = &\frac12 R - g_{i \bar \jmath}D_\mu z^i D^\mu \bar z^{\bar \jmath} + {\rm Im} \mathcal{N}_{\Lambda \Sigma}F_{\mu\nu}^{\Lambda}F^{\mu\nu \, \Sigma} + \frac12 {\rm Re}\mathcal{N}_{\Lambda \Sigma}\epsilon^{\mu\nu\rho\sigma} F_{\mu\nu}^{\Lambda} F_{\rho\sigma}^{\Sigma},
\end{aligned}
\end{equation}
where $F_{\mu\nu}^\Lambda$ are the field strengths of the vectors.
Notice that the matrix ${\rm Im}\mathcal{N}_{\Lambda \Sigma}$ is negative definite, as it can be deduced from the kinetic term of the vector fields.

The gauged theory can be constructed as a deformation of the ungauged one, but we do not discuss it here. The gauging is nevertheless necessary to generate a non-trivial potential for the scalars.

For the purposes of the present review, it is important to discuss symplectic transformations in some more detail. This is because modular transformations will necessarily be a subset of them. The action of a general Sp$(2n_V+2,\mathbb{R})$ matrix on the symplectic sections is
\begin{equation}
\label{eq:symplrotO}
\left(\begin{array}{c}
\tilde{X}^\Lambda\\
\tilde{F}_\Lambda
\end{array}
\right) = \mathcal{V} 
\left(\begin{array}{c}
X^\Lambda\\
{F}_\Lambda
\end{array}
\right), \quad \text{with}\quad
\mathcal{V} =\left(
\begin{array}{cc}
A & B \\
C & D
\end{array}
\right) \in Sp(2n_V+2, \mathbb{R}),
\end{equation}
 or equivalently
\begin{align}
\tilde X^\Lambda &= {A^\Lambda}_\Sigma X^\Sigma + B^{\Lambda \Sigma}F_\Sigma,\\
\tilde F_\Lambda &= {D_\Lambda}^\Sigma F_\Sigma + C_{\Lambda \Sigma} X^\Sigma,
\end{align}
where the matrices $A$, $B$, $C$, $D$ have to satisfy $A^T D - C^T B = D^T A - B^T C = 1$, $A^T C = C^T A$, $B^TD = D^TB$ in order for $\mathcal{V}$ to be an element of Sp$(2n_V+2,\mathbb{R})$.  
Due to supersymmetry, whenever rotating the scalars one rotates also the vectors. By imposing consistency of definition of the rotated vectors, one finds that the gauge kinetic matrix transforms under $Sp(2n_V+2, \mathbb{R})$ as
\begin{equation}
\tilde{\mathcal{N}}_{\Lambda \Sigma} = ({D_\Lambda}^\Gamma \mathcal{N}_{\Gamma \Delta} + C_{\Lambda \Delta}){((A+B\mathcal{N})^{-1})^\Delta}_\Sigma
\end{equation}
which is morally a generalization of the basic SL$(2,\mathbb{R})$ transformation. It is instructive to distinguish among three different kinds of symplectic transformations. 
First, there are classical target-space duality transformations which are symmetries of the tree level action and are given by $C=B=0$, $D^T = A^{-1}$. 
Second, there are semiclassical transformations, $C=0$, $D^T = A^{-1}$, which are not leaving the action invariant in general. 
Third, there are electro-magnetic dualities, $A=D=0$, $C^T = - B^{-1}$, which are such that $\tilde{\mathcal{N}} = - C \mathcal{N}^{-1}C^T$; Since in heterotic compactifications they lead to an inversion of the dilaton $S$, they are called S-dualities. In section \ref{sec:N=2} we review explicit examples of all these kinds of transformations.

\bibliography{references}  
\bibliographystyle{unsrt}

\end{document}